\documentclass[aps,amsmath,superscriptaddress,amssymb,nopacs,nofootinbib,twocolumn,prl]{revtex4-2}

\usepackage[usenames,dvips]{color}
\usepackage{graphicx}
\usepackage{dcolumn}
\usepackage{bm}
\usepackage{mathtools}
\usepackage{epstopdf}
\usepackage{esint}
\usepackage{xcolor}
\usepackage{dsfont}
\usepackage[outline]{contour}
\usepackage{nicefrac}
\usepackage{multirow}

\usepackage{soul}

\usepackage[colorlinks=true,citecolor=magenta,linkcolor=magenta,urlcolor=magenta]{hyperref}

\newcommand{\bea}{\begin{eqnarray}}
\newcommand{\eea}{\end{eqnarray}}
\newcommand{\bt}{\textbf}

\newcommand{\ph}{\phantom{.}}

\newcommand{\noi}{\noindent}
\newcommand{\no}{\nonumber}

\begin{document}

\def\v#1{{\bf #1}}

\title{Imprinting Ground State Chirality on Adatom Spins}

\author{Yun-Peng Huang}
\email{huangyunpeng@iphy.ac.cn}
\affiliation{Beijing National Laboratory for Condensed Matter Physics, and Institute of Physics, Chinese Academy of Sciences, Beijing 100190, China}

\author{Panagiotis Kotetes}
\email{kotetes@itp.ac.cn}
\affiliation{CAS Key Laboratory of Theoretical Physics, Institute of Theoretical Physics, Chinese Academy of Sciences, Beijing 100190, China}

\vskip 1cm

\begin{abstract}
{We propose an alternative experimental protocol for the detection of doped Chern insulators and chiral superconductors. Our approach relies on coupling the target chiral system to adatom spins. Due to the substrate chirality, the adatom spins are expected to order in a noncoplanar configuration with a nonzero spin chirality. Here, we obtain concrete results for chiral substrates which are inva\-riant under arbitrary spin rotations, and are coupled to three adatoms carrying classical moments. By exploring all the accessible magnetic ground states, we identify the regimes in which nonzero spin chirality is induced on the adatom complex. We apply our method to valley-polarized bilayer graphene and $d+id$ superconductors, and find qualitatively different ground state diagrams. Our analysis shows that the adatom spin chirality fully encodes the properties of the substrate chirality.}
\end{abstract}

\maketitle

\textit{\bt{Introduction} $-$} While the unambiguous identification of two-dimensional topological systems   which simultaneously violate parity and time-reversal symmetries~\cite{LeggettRev,FradkinBook,VolovikBook} is highly de\-si\-ra\-ble, it yet remains a formidable task. There exist two main experimental strategies which in principle allow us to detect these also-called chiral systems, i.e., either to observe their topologically protected chiral edge modes, or, to diag\-no\-se the nonzero chirality arising in the bulk. In the former approach, one is typically required to carry out quantum transport experiments~\cite{HgTe,QAHEexp,HeRetracted}, which apart from being extremely challenging to conduct, they may also be inconclusive due to the strong dependence of the measured quantities on the boundary conditions and the possible presence of disorder~\cite{Kayyalha}. On the other hand, the detection of the ground state chirality usually relies on circular-polarization-sensitive responses~\cite{NiuQian,KotetesBook}, such as the polar Kerr effect (PKE)~\cite{PKE}. In spite of the high resolution which is found in state-of-the-art PKE experiments~\cite{KapitulnikRev}, the precise microscopic origin of the observed PKE signal is not always identifiable. A well-known example here is Sr$_2$RuO$_4$, for which a PKE was observed early on~\cite{PKESrRu2O4}, but the type of the pairing gap which breaks time reversal-symmetry still remains unsettled~\cite{MaenoRecentRev}.

\begin{figure}[t!]
\centering
\includegraphics[width=\columnwidth]{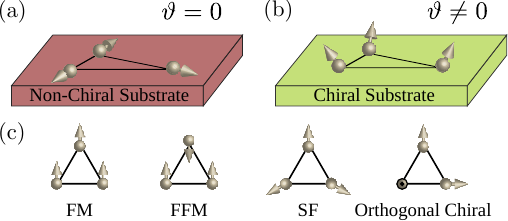}
\caption{Minimal three-adatom setup for the detection of substrate chirality. For a substrate with spin rotational inva\-riance, the adatom spins order in a coplanar fashion when the chirality is zero, as depicted in (a). In contrast, a nonzero chirality promotes a noncoplanar chiral magnetic pattern, see (b). (c) shows the possible magnetic ground states. A non-chiral substrate allows for ferromagnetic (FM), frustrated ferromagnetic (FFM), and spin flux (SF) ground states. When the chirality is switched on, the SF state adiabatically evolves into a noncoplanar chiral state. When chiral spin interactions dominate, the spins order in an orthogonal chiral fashion.}
\label{fig:Figure1}
\end{figure}

In this Letter, we bring forward an alternative method to detect the ground state chirality. We propose to couple the target chiral system to adatom spins, and mo\-ni\-tor the type of ground state which gets stabilized due to the coupling of the spins to the substrate. The minimal configuration which can encode a nonzero ground state chirality is three spins, and is schematically depicted in Fig.~\ref{fig:Figure1}. Here, the adatoms are modeled as point-like objects whose magnetic moments are treated as classical spins. The key observation underlying our detection proposal is that, when the substrate is chiral, the ground state of the spins is expected to possess a nonzero spin chirality $\chi_{\cal C}$~\cite{Villain,Miyashita,WenWilczekZee}. The latter is a measure of the degree of the noncoplanarity of the adatom spin configuration. Hence, the detection of spin chirality promises to provide an indirect signature of the substrate chirality.

We draw concrete conclusions for chiral substrates which are invariant under arbitrary spin rotations. By integrating out the substrate electrons, we obtain an effective functional for the adatom spins which consists of the usual Ruderman-Kittel-Kasuya-Yosida (RKKY) two-spin interaction~\cite{RKKY}, along with a chiral three-spin interaction~\cite{WenWilczekZee,FradkinBook} which encodes the substrate chirality. By identifying all the possible ground states of the spins, we conclude that, when the usual RKKY term favors ferromag\-netic ordering, a nonzero spin chirality can be induced on the spins discontinuously, and only after the strength of the chiral spin interaction reaches a certain threshold. In contrast, when the RKKY term favors antiferromagnetism, a nonzero spin chirality can be induced continuously on the adatom spins.   We make predictions for valley polarized bilayer graphene and $d+id$ superconductors. We find that these systems exhibit qua\-li\-ta\-ti\-ve\-ly dif\-ferent Frie\-del oscillations and spin chiralities.

\textit{\bt{Representative Model Hamiltonian} $-$} We consider the single-particle two-band Hamiltonian $\hat{H}(\bm{k})=\bm{d}(\bm{k})\cdot\bm{\tau}-d_0\hat{\mathds{1}}_\tau$, which depends on the wave vector $\bm{k}=(k_x,k_y)$, and is identical for both spins. In the above, $\tau_{1,2,3}$ define Pauli matrices and $\hat{\mathds{1}}_\tau$ the respective identity matrix. From now on, we omit writing the identity matrix to simplify the notation. Here, $d_0$ defines a generalized chemical potential while we choose $\bm{d}(\bm{k})$ as:
\begin{align}
\bm{d}(\bm{k})=\frac{\Delta}{k_0^2}\Big(k_x^2-k_y^2,\,2 k_xk_y,\,V\big(\bm{k}^2-k_0^2\big)/\Delta\Big)\,.
\label{eq:Hamiltonian}
\end{align}

\noi Depending on the case, the two-state $\tau$ basis can relate to orbital, sublattice, Nambu, or other degrees of freedom.

\textit{\bt{Effective Spin Interactions} $-$} We assume that the substrate electrons couple to a number of $N$ adatom spins $\bm{S}_i$ which are deposited at positions $\bm{R}_i$, by means of the exchange interactions $H_J=J\sum_{i=1}^N\bm{S}_i\cdot\hat{\bm{s}}_i$. Here, $J$ controls the strength of the coupling and $\hat{\bm{s}}_i$ is the electronic spin operator. In the remainder we set $|\bm{S}_i|=1$. By integrating out the substrate electrons, we derive an effective energy functional $E_{\rm spins}$ which contains the multi-spin interactions. For this purpose, we assume that the exchange coupling is sufficiently small, so that we can neglect possible bound states induced by the adatoms~\cite{BalatskyRev}. Under this condition, we treat the substrate-adatom coupling perturbatively, and consider only the contribution of the itinerant electrons~\cite{MartinBatista,Motome,HayamiMotomeRev}. Since in this work it is crucial to capture the nonzero chirality dictating the substrate, it is important to expand at least up to third order in the coupling $J$. By doing so, we find that $E_{\rm spins}$ is a sum of two parts. The first is the usual RKKY term:
\begin{align}
E_{\rm RKKY}=-\frac{J^{2}}{2}\sum_{i,j}\chi_{ij}\bm{S}_{i}\cdot\bm{S}_j
\label{eq:RKKY}
\end{align}

\noi while the second encodes the three-spin chiral coupling:
\begin{align}
E_{\rm Chiral}=-\frac{J^3}{6}\sum_{i,j,k}\vartheta_{ijk}\bm{S}_{i}\cdot\big(\bm{S}_j\times\bm{S}_k\big).
\label{eq:Chiral}
 \end{align}

The chiral spin interaction has so far been discussed in various contexts. It first appeared in the field of spin li\-quids~\cite{Villain,Miyashita,WenWilczekZee}, which remains up to now an open research topic~\cite{Paramekanti2017,JiangJiang,HuangSheng,Paramekanti2023}. While in these studies the chiral spin interaction is treated phenomenologically, other works have discussed how this emerges in the presence of a magnetic field~\cite{DiptimaSen,Scarola}. More recently, it was emphasized that chiral spin terms can also be engineered in topological systems whose bands carry a nonzero Berry curvature~\cite{DongLevitov,HuangMSC,KotetesMSC}, without any requirement for an external magnetic field.

Our goal is to evaluate the two- and three-spin susceptibilities $\chi_{ij}\equiv\chi(\bm{R}_i,\bm{R}_j)$ and $\vartheta_{ijk}\equiv\vartheta(\bm{R}_i,\bm{R}_j,\bm{R}_k)$. In the zero temperature limit, these are defined as:
\bea
\chi_{ij}&=&\,-\,2\int\frac{d\varepsilon}{2\pi}\,{\rm tr}\left[\hat{G}(\varepsilon,\bm{R}_{ij})\hat{G}(\varepsilon,\bm{R}_{ji})\right],\label{eq:RKKY_Sus}
\\
\vartheta_{ijk}&=&-4\imath\int\frac{d\varepsilon}{2\pi}\,{\rm tr}\Big[\hat{G}(\varepsilon,\bm{R}_{ij})\hat{G}(\varepsilon,\bm{R}_{jk})\hat{G}(\varepsilon,\bm{R}_{ki})\Big],\quad\,\,\label{eq:Chiral_Sus}
\eea

\noi where tr denotes trace in $\tau$ space and $\bm{R}_{ij}=\bm{R}_i-\bm{R}_j$. When $\tau$ labels electron and hole states of a superconductor, a factor of $\nicefrac{1}{2}$ needs to be added to Eqs.~\eqref{eq:RKKY_Sus} and~\eqref{eq:Chiral_Sus} to avoid double counting the electronic degrees of freedom.

We observe that the coefficients $\chi_{ij}$ and $\vartheta_{ijk}$ depend only on the differences of the adatom positions, since the substrate is assumed to be invariant under arbitrary spatial translations. The two susceptibilities are expressed in terms of the coordinate space matrix Green function:
\begin{align}
\hat{G}(\varepsilon,\bm{R})=\int\frac{d\bm{k}}{(2\pi)^2}\,e^{\imath\bm{k}\cdot\bm{R}}\,\hat{G}(\varepsilon,\bm{k}),
\end{align}

\noi which is the Fourier transform of the Euclidean matrix Green function $\hat{G}(\varepsilon,\bm{k})=1/\big[\imath\varepsilon-\hat{H}(\bm{k})\big]$, where $\varepsilon\in(-\infty,+\infty)$ defines the imaginary energy. We provide technical details in our supplemental material~\cite{SM}.

{\textit{\bt{Magnetic Ground States of Three Adatoms $-$}}}
In order to experimentally detect the nonzero chirality of the substrate, it is crucial to first identify the set of ground states that become accessible for the minimal setup which consists of three adatoms at positions $\bm{R}_{1,2,3}$ carrying magnetic moments $\bm{S}_{1,2,3}$, respectively. By combining the RKKY and chiral contributions of Eqs.~\eqref{eq:RKKY} and~\eqref{eq:Chiral}, the corresponding spin functional takes the form:
\bea
E_{\rm spins}&=&-J^2\big(\chi_{12}\bm{S}_1\cdot\bm{S}_2+\chi_{13}\bm{S}_1\cdot\bm{S}_3+\chi_{23}\bm{S}_2\cdot\bm{S}_3\big)\no\\
&&-J^3\vartheta\bm{S}_{1}\cdot\big(\bm{S}_{2}\times\bm{S}_{3}\big)\,,
\label{eq:EffectiveSpinHamiltonian}
\eea

\noi where we set  $\vartheta=\vartheta_{123}$. Given the coupling coefficients $\chi_{12},\chi_{13},\chi_{23}$ and $\vartheta$, we extremize $E_{\rm spins}$ and infer all the possible  magnetic ground states. For details see Ref.~\onlinecite{SM}.

By first setting $\vartheta=0$ and allowing for ge\-ne\-ral positions $\bm{R}_{1,2,3}$, we find three types of magnetic extrema: (i) a ferromagnetic (FM) state with all spins aligned, (ii) the here-termed frustrated ferromagnetic (FFM) state where one spin is flipped compared to the FM state, and (iii) the coplanar spin flux (SF) state, where the spins are all pointing mostly inwards or mostly outwards, thus gi\-ving rise to a ``combed" planar spin ``hedgehog". When the adatoms form an equilateral triangle, the spins in the SF state for an ideal ``hedgehog" pattern with the spin flux flowing inwards or outwards being ma\-xi\-mal, since successive spins form an angle of $2\pi/3$, as depicted in Fig.~\ref{fig:Figure1}(c). Straightfoward manipulations show that in the equilateral triangle setup, where $\chi_{12,13,23}\equiv\chi$, the FM state is the ground state for $\chi>0$ with an energy $E_{\rm FM}=-3J^2|\chi|$, while SF becomes stabilized for $\chi<0$ with an energy $E_{\rm SF}=-3J^2|\chi|/2$. As a result, the FFM does not appear in this highly symmetric adatom setup.

When $\vartheta$ is switched on, one finds three ground states once again, i.e., the FM and FFM states discussed above, along with a chiral magnetic (CM) state which possesses a nonzero spin chirality $\chi_{\cal C}=\bm{S}_1\cdot\big(\bm{S}_2\times\bm{S}_3\big)$~\cite{WenWilczekZee}. The CM ground state enjoys maximum spin chirality when the spins are orthogonal to each other, as shown in Fig.~\ref{fig:Figure1}(c). This here-termed orthogonal CM state, is achievable in an equilateral triangle setup when the magniture of the RKKY suscep\-ti\-bi\-li\-ty $|\chi|$ is much smaller than $|J\vartheta|$. A sweet spot where this inequality is automatically satisfied is when $\chi$ vanishes while at the same time $\vartheta$ is nonzero.

We now investigate the competition of the CM state and the collinear ones for the two possible signs of $\chi$. For $\chi>0$ the CM competes with FM. We find that the CM ground state is accessible only for $|J\vartheta|\geq\sqrt{3}\chi$. Remarkably, in the same interval that the CM state is well-defined, it also becomes the ground state, with its energy being exactly equal to $E_{\rm FM}=-3J^2\chi$ at $|J\vartheta|=\sqrt{3}\chi$, and further decreasing for $|J\vartheta|/\chi>\sqrt{3}$. Hence, for $\chi>0$, a threshold value $|J\vartheta|/\chi=\sqrt{3}$ is required to be reached so to induce spin chirality on the adatoms.

In stark contrast, when $\chi=-|\chi|<0$, the CM solution is always accessible. As a matter of fact, in this parameter regime, the limit $\vartheta\rightarrow0$ is well-defined and the CM state energy becocomes equal to $-3J^2|\chi|/2$. Hence, the CM state evolves into the SF state, thus, implying that SF and CM states are adiabatically connected and constitute the ground states of the adatom complex for $\chi<0$. Also here, the FFM state does not get stabilized.

It is important here to point out that the chiral spin interaction acts as a ``catalyst" for noncoplanar phases carrying a nonzero spin chirality. We find that when the ground state is nonchiral and the FM state is stabilized, switching on $\vartheta$ does not immediately induce a nonzero spin chi\-ra\-li\-ty on the adatoms. However, when starting from the SF state, chirality is immediately sourced by a nonzero $\vartheta$. An analogous phenomenon was pre\-viou\-sly discussed for bulk two-dimensional iti\-ne\-rant magnets in prior works of ours~\cite{HuangMSC,KotetesMSC}. There, we showed that switching on a chiral interaction on top of the so-called spin-vortex~\cite{Lorenzana,Christensen} or spin-whirl crystal phases~\cite{Christensen} induces adiabatically two distinct types of spin-skyrmion crystals phases. The adiabatic evolution of nonskyrmion phases into skyrmionic ones is possible, because the spin-vortex and -whirl textures carry topological charges~\cite{Christensen}.

A similar situation takes place here for the system of the three adatoms. For $\vartheta=0$, the SF state is the only one which carries a nontrivial topological charge, which gives rise to a nonzero vector spin chirality~\cite{HayamiVectorChirality,HuangSen}. This is precisely the reason why the SF ground state enables the induction of scalar spin chirality even upon adding an infi\-ni\-te\-si\-mal\-ly weak chiral interaction. However, this is not possible when starting from the FM state due its to\-po\-lo\-gi\-cal incompatibility with the CM state. Therefore, here, spin chirality can be only induced discontinuously.

\textit{\bt{Valley-Polarized Bilayer Graphene $-$}} A prominent realization of  Chern insulators~\cite{VolovikBook,HuangMSC,FuTCI,BalatskyChiraldwave,Raghu,Tewari,KotetesMeissner,ChuanweiZhang,TiltedHillNernst,Venderbos}, can be found in bilayer graphene. In the low-energy regime, bilayer graphene can be effectively modeled by means of a two band
Hamiltonian per valley~\cite{Fal'ko2006}. Each valley Hamiltonian takes the general form shown in Eq.~\eqref{eq:Hamiltonian}. In the present case, the $\tau$ basis is spanned by electrons living on sublattice A of one layer and electrons of sublattice B of the other layer. Note that A and B correspond to the two sublattices comprising the honeycomb lattice. The mapping of Eq.~\eqref{eq:Hamiltonian} to the model of  Ref.~\onlinecite{Fal'ko2006} is achieved by setting $\Delta=\gamma_1/2$ and $k_0=\gamma_1/\sqrt{2}\hbar\upsilon_{\rm g}$. Here, $\gamma_1\simeq390~{\rm meV}$ is the energy scale of the do\-mi\-nant interlayer coupling, $\upsilon_{\rm g}\simeq8\cdot10^5\,{\rm m/s}$ is the Dirac ve\-lo\-ci\-ty of graphene, and $\hbar$ defines the reduced Planck constant. Our analysis neglects the effects of trigonal war\-ping, since here we restrict to the relatively high-energy regime. In this limit, the Fermi surface consists of a single big pocket~\cite{Fal'ko2006}. The spin-chirality phenomena predicted in this work become accessible only when a net valley polarization is present~\cite{ValleyContrasting}. For instance, this can take place when the parameter $V$ differs for the two valleys. The variable $V$ corresponds to the half-difference of the electrochemical potentials defined for the two layers and is typically engineered by gating the bilayer~\cite{GateValley1,GateValley2,GateValley3}.

By means of straightfoward analytical calculations and a few standard approximations~\cite{SM}, we recover the known expression for the suscep\-ti\-bi\-li\-ty of a given valley:
\begin{equation}
\chi_{\rm valley}= \frac{4\nu_Fk_F^2}{\pi}\frac{\sin\left(2k_{F}R\right)}{\left(2k_{F}R\right)^2}\,,
\end{equation}

\noi which was previously found in Ref.~\onlinecite{DasSarma2008}. Notably, this result is substantially modified away from the low-energy limit~\cite{CatroNeto2009,Satpathy2011,Parhizgar,DasSarma2017}. In the above, we defined the interadatom distance $R>0$, the Fermi wavenumber $k_F$, and the density of states $\nu_F$ at the Fermi level which is set by the che\-mi\-cal po\-ten\-tial $d_0=E_F$. From Ref.~\onlinecite{Fal'ko2006} we find that:
\bea
\frac{k_F}{k_0}&=&\sqrt{\frac{V^2+\sqrt{\big(\Delta^2+V^2\big)E_F^2-\big(V\Delta \big)^2}}{V^2+\Delta^2}}\,,\\
\nu_F&=&\frac{k_0^2}{4\pi}\frac{E_F}{ \sqrt{(\Delta^2+V^2)E_F^2-(V\Delta)^2}}.
\eea

In the same spirit, we now obtain the chiral suscep\-ti\-bi\-li\-ty for the case of an equilateral triangle with an adatom distance which is denoted $R$. For a single valley we have:
\begin{align}
\vartheta_{\rm valley} =\frac{3!(\nu_Fk_F)^2}{\sqrt{8\pi/3}}\left(1-\frac{k_0^2}{k_F^2}\right)\frac{k_F^6}{k_0^6}\frac{V\Delta^2}{E_F^3}\frac{\sin\big(3k_FR-\frac{3\pi}{4}\big)}{(k_FR)^{5/2}}.
\label{eq:ThetaValley}
\end{align}

\noi We remark that the result for a general triangular confi\-gu\-ra\-tion is presented in Ref.~\onlinecite{SM}. From the above, we observe that the chiral susceptibility is odd under $V\mapsto-V$. Based on this property we indeed confirm that for time-reversal-symmetric bilayer graphene the two valleys carry opposite chirality which cancels the total $\vartheta$. Hence, the outcome for the induced adatom chirality strongly depends on the degree of valley polarization which is ma\-xi\-mi\-zed when the two valleys feel identical gaps $V$.

\begin{figure}[t!]
\centering\includegraphics[width=\columnwidth]{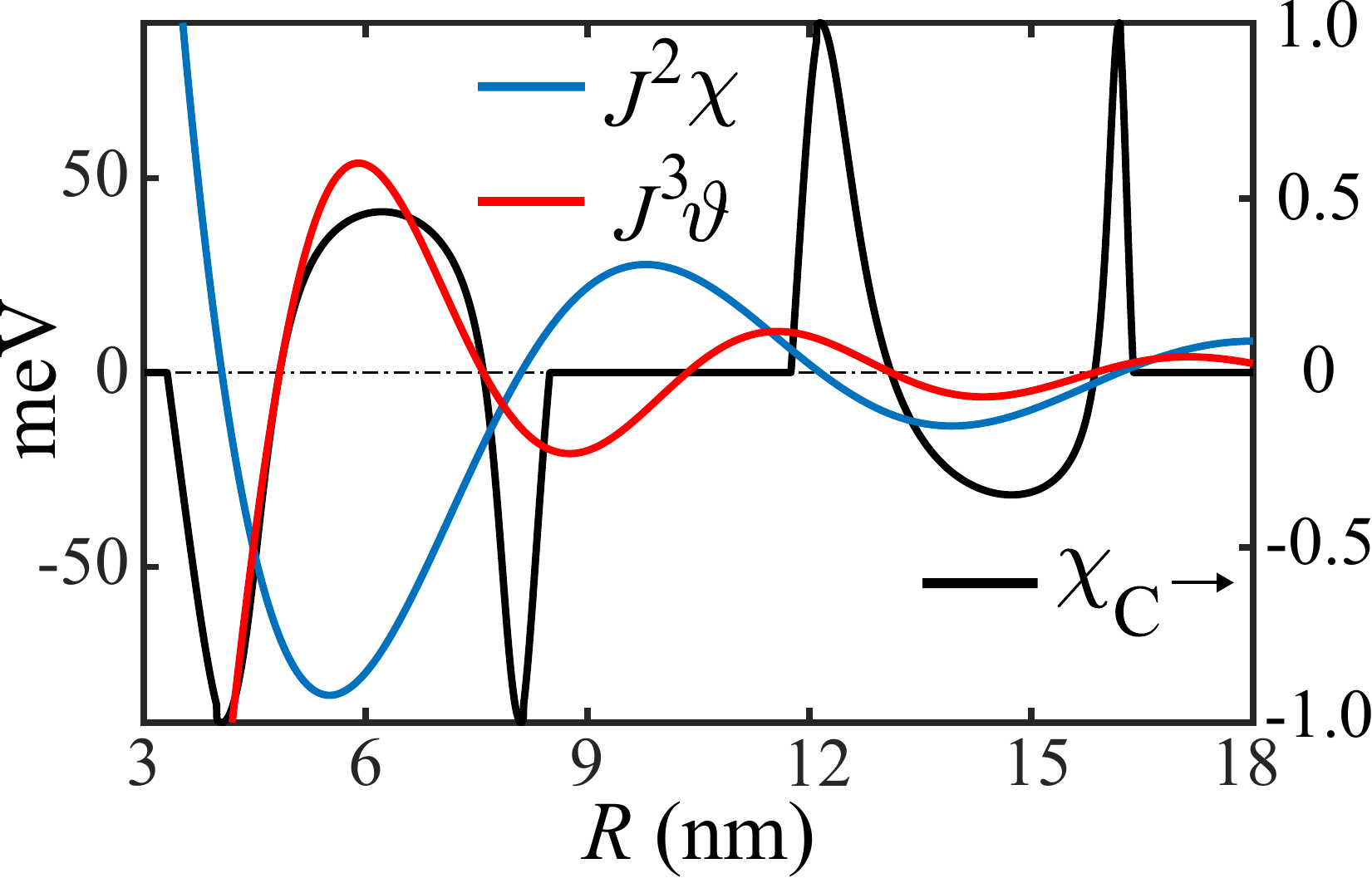}
\caption{RKKY and chiral coefficients $\chi$ and $\vartheta$ as a function of the interadatom distance $R$, for an equilateral triangle of three adatom spins coupled to bilayer graphene. By choo\-sing $V_{\bm{K}}=90\,{\rm meV}$ and $V_{\bm{K}'}=-10\,{\rm meV}$, we obtain a valley polarization degree $\rm{VP}=0.8$. We further set $E_F=110\,{\rm meV}$ and $Jk_F^2=0.9\,\rm{eV}$. The spin chirality $\chi_{\cal C}$ is nonzero only in the CM state, it becomes maximized when $\chi=0$, while its sign tracks the sign of $\vartheta$. For $\vartheta=0$ and $\chi>0$ ($\chi<0$) the spins order in the FM (SF) configuration with $\chi_{\cal C}=0$.}
\label{fig:Figure2}
\end{figure}

In Fig.~\ref{fig:Figure2}, we show representative results for an equi\-la\-te\-ral triangular spin setup, when accoun\-ting for both $\bm{K}$ and $\bm{K}'$ valleys. We define the valley polarization degree as ${\rm VP}=\big|V_{\bm{K}}+V_{\bm{K}'}\big|/\big(|V_{\bm{K}}|+|V_{\bm{K}'}|\big)$ and show results for ${\rm VP}=0.8$. We observe that the ground state mainly alternates between the FM and CM states, while the SF state is stabilized when $\vartheta=0$ and $\chi<0$. The spin chirality is nonzero in the CM state with maxima appearing for $\chi=0$. Notably, these maxima are easily reachable since the two susceptibilities are dictated by a different sinusoidal dependence on $R$. When instead $|\chi|\sim|J\vartheta|$, we find moderate values for the spin chirality.

\textit{\bt{Application to chiral superconductors $-$}} Our model in Eq.~\eqref{eq:Hamiltonian} also applies to $d+id$ spin-singlet superconductors~\cite{LeggettRev,VolovikBook,Laughlin} after setting $d_0=0$. In contrast to the case of bilayer graphene, here the Fermi level is set by $V=E_F$. The two-level $\tau$ Hilbert space is spanned by electron and hole states, which describe Cooper pairs with a gap $\Delta>0$. Also here we follow the same approach~\cite{SM}. Specifically, we employ the quasi-classical approximation $E_F\gg\Delta$ and further restrict to si\-tuations where the largest interadatom distance is much smaller than the superconduc\-ting co\-he\-ren\-ce length $\xi_{\rm sc}=\hbar\upsilon_F/\Delta$, where $\upsilon_F$ is the Fermi velocity in the normal phase. Our calculations~\cite{SM} recover the known result~\cite{Aristov1997}:
\begin{align}
\chi_{d+id}=\frac{4\nu_Fk_F^2}{\pi}\left[\frac{\sin\big(2k_FR\big)}{\big(2k_FR\big)^2}-\frac{\pi}{4}\frac{1}{k_F^2\xi_{\rm sc}R}\right],
\end{align}

\begin{figure}[t!]
\centering
\includegraphics[width=\columnwidth]{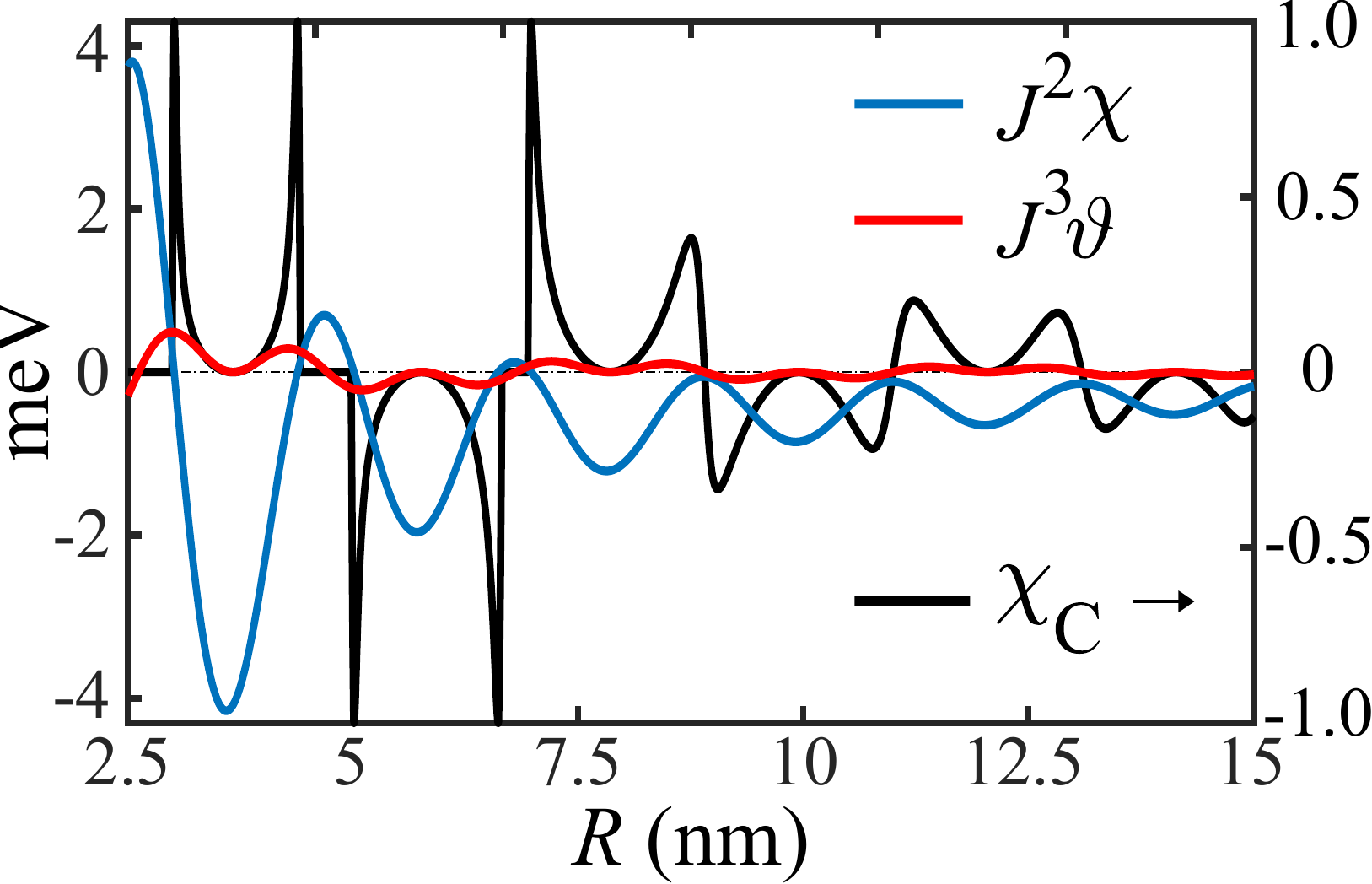}
\caption{Same as in Fig.~\ref{fig:Figure2}, but for spins coupled to a $d+id$ superconductor. Here, we set $\Delta=8\,\rm{meV}$, $\xi_{\rm sc}=50\,\rm{nm}$, $E_F=300\,\rm{meV}$, and $Jk_F^2=1\,\rm{eV}$.  The presence of the spin-singlet pairing gap strongly affects the sign and Friedel oscillation period of the two coefficients, thus promoting the CM state.}
\label{fig:Figure3}
\end{figure}

\noi where in the present case $k_F$ coincides with $k_0$, and the normal phase density of states at the Fermi level is defined as $\nu_F=k_F/2\pi\xi\Delta$. The first term corresponds to the familiar RKKY interaction obtained when the substrate is a good metal. On the other hand, the second term appears only in the presence of the spin-singlet pai\-ring gap~\cite{Aristov1997}. As it has been previously pointed out~\cite{Yao2014,Schecter2016,Schecter2018}, this second term can dominate the RKKY suscepti\-bi\-li\-ty and favors antiferromagnetic ordering. This behavior stems from the spin-singlet character of the pairing and tends to promote the CM over the FM ground state.

Under the same approximations, we derive the coefficient of the chiral spin interaction, which reads as:
\begin{align}
\vartheta_{d+id}=-(6\pi)^{\frac{3}{2}} \nu_F^3\Delta\frac{\cos^2\left(k_FR-\frac{\pi}{4}\right)\sin\left(k_FR-\frac{\pi}{4}\right)}{\left(k_FR\right)^{3/2}}\,.
\label{Eq:SC_Theta}
\end{align}

\noi We observe that the structure of $\vartheta_{d+id}$ differs substantially to the one in Eq.~\eqref{eq:ThetaValley}. This concerns both the period for the Friedel oscillations, as well as the dependence on $k_FR$. These differences can be attributed to the Cooper-pairing mecha\-nism which is responsible here for generating the ground state chirality of the substrate.

Figure~\ref{fig:Figure3} depicts the evolution of the two suscepti\-bi\-li\-ties along with the arising spin chirality, upon varying the interadatom distance $R$, in the case of an equilateral triangular configuration. Remarkably, aside from the anticipated spin chirality peaks centered at $\chi=0$, the spin-singlet pairing promotes the CM state, and renders $\chi_{\cal C}$ nonzero in a substantial portion of the parameter space.

\textit{\bt{Discussion} $-$} From the above, we conclude that our chirality detection protocol is broadly applicable, it can differentiate between normal and superconducting systems and, most importantly, it fully encodes the winding of the Hamiltonian in the Friedel oscillations of the chiral susceptibility. Our approach appears expe\-ri\-men\-tal\-ly feasible, since adatom spins can be addressed using scanning tunneling microscopy~\cite{WiesendangerRev}. Evenmore, the substrate-adatom coupling can be electrically controlled~\cite{Nafday2013,Nafday2016,Killi2011}, while the geometrical details of the setup are also adjustable, thus providing additional control knobs.

Aside from valley-polarized bilayer graphene, our approach appears applicable to magic-angle twisted bilayer graphene, which is known to develop spontaneous Chern phases~\cite{Serlin,Nuckolls}. Another possibility is the charge-ordered kagom\'e superconductors~\cite{ChiralKagome,KagomeCascade}. The observed charge-order is theoretically proposed to be a chiral phase~\cite{KagomeFluxPhase,KagomeFluxPhaseClassi,NandkishoreKagome} akin to the one proposed by Haldane~\cite{Haldane}. Note that similar to its cousin, i.e., the quantum anomalous Hall state~\cite{QAHE,LawQAHE}, also the charge Chern insulator can be employed for engineering topological superconductivity~\cite{MajoranaRacetrack}. Apart from doped Chern insulators, our protocol can be applied to candidate chiral $d+id$ superconductors, such as, SrPtAs~\cite{SrPtAsExp1,SrPtAsExp2,SrPtAsFischer}, cuprate bilayers~\cite{Franz} based on Bi-2212 monolayers~\cite{Yu:Bi-2212}, and intrinsically superconducting graphene when tuned near van Hove singularities~\cite{BlackSchaffer,Pathak,Nandkishore,Kiesel}.

\section*{Note Added}

At the final stage of completion of this manuscript, we became aware of the preprint in Ref.~\cite{LevitovSpinChirality} which has some overlap with our work.

\section*{Acknowledgements}

PK acknowledges funding from the National Na\-tu\-ral Science Foundation of China (Grant No.~12250610194).

\appendix

\begin{widetext}

\section{{\huge
Supplemental Material}}

\section{{\large Imprinting Ground State Chirality on Adatom Spins}}

\subsection{Yun-Peng Huang$^1$ and Panagiotis Kotetes$^2$}
\begin{center}
$^1$ Beijing National Laboratory for Condensed Matter Physics, and Institute of Physics,\\ Chinese Academy of Sciences, Beijing 100190, China\\
$^2$ CAS Key Laboratory of Theoretical Physics, Institute of Theoretical Physics,\\ Chinese Academy of Sciences, Beijing 100190, China
\end{center}

\section{{\large A. Magnetic Ground States}}

We base our analysis on the effective chiral Heisenberg spin model of the main text. We choose to parametrize the spins of the three adatoms in the following fashion $\bm{S}_1=\big(0,0,1\big)$, $\bm{S}_2=\big(0,\sin\phi,\cos\phi\big)$, and $\bm{S}_3=\big(\sin\omega\sin\eta,\sin\omega\cos\eta,\cos\omega\big)$, where $\phi\in[0,2\pi)$, $\omega\in[0,\pi]$, and $\eta\in[0,2\pi)$. After replacing the spins in terms of the angles $(\phi,\omega,\eta)$, we find the energy functional:
\bea
E_{\rm spins}=-J^2\left[\chi_{12}\cos\phi+\chi_{13}\cos\omega+\chi_{23}\cos\phi\cos\omega+\sin\phi\sin\omega\big(\chi_{23}\cos\eta-J\vartheta\sin\eta\big)\right].
\eea

\noi We proceed by extremizing the functional. We first extremize with respect to $\eta$, i.e., $\partial E_{\rm spins}/\partial\eta=0$, and find:
\begin{align}
\sin\phi\sin\omega\big(\chi_{23}\sin\eta+J\vartheta\cos\eta\big)=0,
\end{align}

\noi which leads to the following possibilities:
\begin{itemize}
\item $\phi=\{0,\pi\}$. In these cases $\bm{S}_1$ and $\bm{S}_2$ are collinear. Further extremization with respect to $\omega$ yields the two subcases $\omega=\{0,\pi\}$. Therefore, in this case all the spins are collinear and there exist two possible ground states, the ferromagnetic (FM) and the here-termed frustrated FM (FFM). In the latter, one of the three spins which are aligned in the FM state flips direction. The energy for the FM alignment is given by $E_{\rm FM}=-J^2\big(\chi_{12}+\chi_{13}+\chi_{23}\big)$. In contrast, for the FFM scenario we find three nonequivalent possibilities when $\chi_{12}\neq\chi_{13}\neq\chi_{23}$. Hence, depending on which spin is flipped, we find $E_{\rm FFM}^{\uparrow\uparrow\downarrow}=-J^2\big(\chi_{12}-\chi_{13}-\chi_{23}\big)$, $E_{\rm FFM}^{\uparrow\downarrow\uparrow}=-J^2\big(\chi_{13}-\chi_{12}-\chi_{23}\big)$, and $E_{\rm FFM}^{\downarrow\uparrow\uparrow}=-J^2\big(\chi_{23}-\chi_{12}-\chi_{13}\big)$.

\item $\omega=\{0,\pi\}$. Due to the arising symmetry of the energy functional under the exchange $\phi\leftrightarrow\omega$, this and the previous case lead to the same ground states.

\item $\vartheta\neq0$ and $\tan\eta=-J\vartheta/\chi_{23}$. This corresponds to the noncoplanar chiral ground state. By making use of this relation, we find that the energy now becomes:
\bea
E_{\rm spins}=-J^2\left[\chi_{12}\cos\phi+\chi_{13}\cos\omega+\chi_{23}\cos\phi\cos\omega-{\rm sgn}(J\vartheta\sin\eta)\sqrt{\chi_{23}^2+(J\vartheta)^2}\sin\phi\sin\omega\right].
\eea

\noi We proceed by further extremizing with respect to $\omega$ and find the condition:
\begin{align}
\tan\omega=-\frac{{\rm sgn}( J\vartheta\sin\eta)\sqrt{\chi_{23}^2+(J\vartheta)^2}\sin\phi}{\chi_{13}+\chi_{23}\cos\phi}\,.
\end{align}

\noi By making use of the above relation, we obtain the following form of the energy functional at this extremum:
\bea
E_{\rm spins}=-J^2\left\{\chi_{12}\cos\phi-{\rm sgn}(J\vartheta\sin\eta \sin\phi)\sqrt{\big(\chi_{13}+\chi_{23}\cos\phi\big)^2+\big[\chi_{23}^2+(J\vartheta)^2\big]\sin^2\phi}\right\}.
\eea

\noi We remind the reader that since $\phi\neq\{0,\pi\}$ the function $\sin\phi$ is nonzero. Thus, we can proceed with the extremization in terms of $\phi$ by separately considering the two cases depending on the value of ${\rm sgn}(\sin\phi)$. By extremizing with respect to $\phi$ in each case, we find:
\begin{align}
\chi_{12}={\rm sgn}( J\vartheta\sin\eta\sin\phi)\frac{\chi_{13}\chi_{23}-(J\vartheta)^2\cos\phi}{\sqrt{\big(\chi_{13}+\chi_{23}\cos\phi\big)^2+\big[\chi_{23}^2+(J\vartheta)^2\big]\sin^2\phi}}\,,
\end{align}

\noi and after solving for $\cos\phi$, we find the two possible solutions:
\begin{align}
\frac{\big[\chi_{12}^2+(J\vartheta)^2\big]\cos\phi-\chi_{13}\chi_{23}}{\chi_{12}}=\frac{\chi_{12}\chi_{13}\chi_{23}\pm \sqrt{\big[\chi_{12}^2+(J\vartheta)^2\big]\big[\chi_{13}^2+(J\vartheta)^2\big]\big[\chi_{23}^2+(J\vartheta)^2\big]}}{(J\vartheta)^2}\,,
\end{align}

\noi which results in the ground state energy of the chiral magnetic state:
\bea
E_{\rm CM}=-\frac{|\chi_{12}\chi_{13}\chi_{23}|}{\vartheta^2}\left\{{\rm sgn}(\chi_{12}\chi_{13}\chi_{23})\pm \sqrt{\left[1+\left(\frac{J\vartheta}{\chi_{12}}\right)^2\right]\left[1+\left(\frac{J\vartheta}{\chi_{13}}\right)^2\right]\left[1+\left(\frac{J\vartheta}{\chi_{23}}\right)^2\right]}\right\}.
\eea

\noi The above sign ambiguity is resolved by choosing the solution for $\cos\phi$ which minimizes the energy. Since the term under the square root is larger than one, we find that independent of the sign of $\chi_{12}\chi_{13}\chi_{23}$ the ``$+$" solution minimizes the energy. Hence, one can now trace the steps backwards and identify the various parameters.

\item $\vartheta=0$ and $\eta=\{0,\pi\}$. This state corresponds to the coplanar here-termed spin flux (SF) state. Given these conditions the energy functional takes the form below:
\bea
E_{\rm spins}=-J^2\left(\chi_{12}\cos\phi+\chi_{13}\cos\omega+\chi_{23}\cos\phi\cos\omega+e^{i\eta}\chi_{23}\sin\phi\sin\omega\right).
\eea

\noi Extremizing with respect to $\omega$ leads to the condition:
\begin{align}
\tan\omega=\frac{e^{i\eta}\chi_{23}\sin\phi}{\chi_{13}+\chi_{23}\cos\phi},
\end{align}

\noi which allows us to write the energy of this extremum as follows:
\bea
E_{\rm spins}=-J^2\left[\chi_{12}\cos\phi+e^{i\eta}{\rm sgn}\big(\chi_{23}\sin\phi\big)\sqrt{(\chi_{13}+\chi_{23}\cos\phi)^2+(\chi_{23}\sin\phi)^2}\right].
\eea

\noi In the same spirit with the process followed in the previous paragraph for the CM state, here extremizing yields:
\begin{align}
\chi_{12}=-e^{i\eta}{\rm sgn}\big(\chi_{23}\sin\phi\big)\frac{\chi_{13}\chi_{23}}{\sqrt{(\chi_{13}+\chi_{23}\cos\phi)^2+(\chi_{23}\sin\phi)^2}},
\end{align}

\noi as well as the solution:
\begin{align}
\chi_{12}^2\cos\phi-\chi_{13}\chi_{23}=-\frac{(\chi_{12}\chi_{13})^2+(\chi_{12}\chi_{23})^2+(\chi_{13}\chi_{23})^2}{2\chi_{13}\chi_{23}}.
\end{align}

\noi Using the above, we find the energy of this ground state:
\begin{align}
E_{\rm SF}=J^2\frac{(\chi_{12}\chi_{13})^2+(\chi_{12}\chi_{23})^2+(\chi_{13}\chi_{23})^2}{2\chi_{12}\chi_{13}\chi_{23}}.
\end{align}

\end{itemize}

Before proceeding, we examine how the above results for the four distinct magnetic states simplify when the adatoms form an equilater triangle, where $\chi_{12,13,23}\equiv\chi$.

\begin{itemize}

\item FM state. The energy becomes $E_{\rm FM}=-3J^2\chi$.

\item FFM state. The energy in this case is given by $E_{\rm FFM}=J^2\chi$.

\item SF state. Here, we find $E_{\rm SF}=3J^2\chi/2$. When $\vartheta=0$ and $\chi<0$, the SF is stabilized as the ground state. By choosing $\eta=\pi$ with no loss of generality, we find $\phi=\omega=2\pi/3$. Thus, successive spins form an angle of $2\pi/3$.

\item CM state. In this case the energy becomes:
\begin{align}
E_{\rm CM}=-J^2\chi\,\frac{1+{\rm sgn}(\chi)\sqrt{1+(J\vartheta/\chi)^2}^3}{(J\vartheta/\chi)^2}\,,
\end{align}

\noi while $\cos\phi$ is to be determined by the following relation:
\begin{align}
\cos\phi=\frac{1}{1+(J\vartheta/\chi)^2}\left[1+\frac{1+{\rm sgn}(\chi) \sqrt{1+(J\vartheta/\chi)^2}^3}{(J\vartheta/\chi)^2}\right].
\end{align}

\noi We investigate the competition of the CM state and the collinear ones for the two signs of $\chi$.
\begin{itemize}
\item[(i)] $\chi>0$. Here, the CM competes with FM. We find that $\cos\phi$ does not have a solution for $|J\vartheta|<\sqrt{3}\chi$, while $\cos\phi\in[0,1]$ for $|J\vartheta|\geq\sqrt{3}\chi$. Note that $\cos\phi=1$ ($\cos\phi=0$) for $|J\vartheta|=\sqrt{3}\chi$ ($|J\vartheta|/\chi\rightarrow\infty$). Interestingly, in the same interval that the CM state is well-defined, it also becomes the ground state, with its energy being exactly equal to $E_{\rm FM}=-3J^2\chi$ at $|J\vartheta|=\sqrt{3}\chi$, and further decreasing upon increasing $|J\vartheta|/\chi$ beyond the value~$\sqrt{3}$. Hence, we observe that a critical strength of $|J\vartheta|/\chi$ is required to induce spin chirality on the adatom spins.

\item[(ii)] $\chi=-|\chi|<0$. In this parameter regime, there always exist a CM solution. Specifically, we find that $\cos\phi\in[-1/2,0)$, with the minimum (maximum) value obtained for $|J\vartheta/\chi|=0$ ($|J\vartheta/\chi|\rightarrow\infty$). Quite remarkably, for $\chi<0$ the limit $\vartheta\rightarrow0$ is well-defined and the CM state energy becocomes equal to $-3J^2|\chi|/2$. Hence, the CM state evolves into the SF state, thus, implying that SF and CM states are adiabatically connected and constitute the ground states of the adatom complex for $\chi<0$. Finally, we conclude that the FFM state never becomes the ground state solution.
\end{itemize}

\end{itemize}

\section{{\large B. Calculations for Bilayer Graphene}}

In this section, we provide further details on obtaining the RKKY and chiral susceptibilities via the Euclidean Green function approach.

\subsection{I. Model Hamiltonian}

We formally bring the BG Hamiltonian $\hat{H}(\bm{k})$ into a diagonal form by expressing it as $\hat{H}(\bm{k})=\sum_{\alpha=\pm}\varepsilon_\alpha(\bm{k})\hat{P}_\alpha(\bm{k})$, where we have introduced the projectors $\hat{P}_\alpha(\bm{k})=\big[\hat{\mathds{1}}+\alpha\hat{\bm{d}}(\bm{k})\cdot\bm{\tau}\big]/2$ and the unit vector $\hat{\bm{d}}(\bm{k})=\bm{d}(\bm{k})/|\bm{d}(\bm{k})|$. These projectors correspond to the energy dispersions $\varepsilon_{\alpha}(\bm{k})=\alpha|\bm{d}(\bm{k})|-\mu\equiv\varepsilon_{\alpha}(k)$, where $k=|\bm{k}|$. Here, we have $|\bm{d}(\bm{k})|^2=\Delta^2(k/k_0)^4+V^2\big[(k/k_0)^2-1\big]^2$. Straightfoward algebra yields that $|\bm{d}(\bm{k})|$ has a local maximum at $k=0$ with $|\bm{d}(\bm{k})|_{\rm local\, max}=|V|$, while it reaches its global minimum for $k=k_0|V|/\sqrt{\Delta^2+V^2}$ with $|\bm{d}(\bm{k})|_{\rm global\, min}=|V\Delta|/\sqrt{\Delta^2+V^2}\equiv E_{\rm min}$. When $|V|\ll|\Delta|$ we find that $E_{\rm min}\rightarrow|V|$ and the two extrema coincide. From the above, we conclude that in order to obtain a Fermi surface, the chemical potential should satisfy $|\mu|>E_{\rm min}$, while the given model is valid in the interval $10^{-2}\leq E_F/|\Delta|\leq2$. This is because $\Delta=\gamma_1/2$ and $k_0=\gamma_1/\sqrt{2}\hbar\upsilon_{\rm g}$, where $\gamma_1=390\,{\rm meV}$ and $\upsilon_{\rm g}\simeq8\cdot10^5\,{\rm m/s}$. In the following, we focus on the case $\mu=E_F>0$, and find that only the $\varepsilon_+(\bm{k})$ band leads to a Fermi surface, with a Fermi wavenumber:
\begin{align}
\left(\frac{k_F}{k_0}\right)^2=\frac{E_{\rm min}}{\sqrt{\Delta^2+V^2}}\left[\left|\frac{V}{\Delta}\right|+\sqrt{\left(\frac{E_F}{E_{\rm min}}\right)^2-1}\right]\,,
\end{align}

\noi since we have $|\Delta|>|V|$. Note that when $|\Delta|\gg|V|$ and $E_F\simeq|\Delta|$, we have $k_F\simeq k_0$. In the remainder we consider that $E_F$ is substantially larger than $|V|$, so that it gives rise to a sufficiently large Fermi pocket which dominates the RKKY and chiral interactions. Hence, under these assumptions, we can completely neglect the gapped dispersion $\varepsilon_-(k)$. In contrast, we keep the dispersion $\varepsilon_+(k)$ and additionally linearize it about the Fermi wavenumber. This procedure yields the result:
\begin{align}
\varepsilon_+(k)=\frac{2k_F}{E_Fk_0^2}\sqrt{(\Delta^2+V^2)E_F^2-(V\Delta)^2}\,\big(k-k_F\big)\,.
\end{align}

\subsection{II. Coordinate-Space Matrix Green Function}

Under the single-dispersion approximation discussed in the previous paragraph, the reciprocal space matrix Green function is expressed as $
\hat{G}(\varepsilon,\bm{k})=\sum_{\alpha=\pm}\big[\imath\varepsilon-\varepsilon_\alpha(k)\big]^{-1}\hat{P}_\alpha(\bm{k})\approx\big[\imath\varepsilon-\varepsilon_+(k)\big]^{-1}\hat{P}_+(\bm{k})$. We now proceed and evaluate the coordinate-space matrix Green function $\hat{G}(\varepsilon,\bm{R})$ by means of a Fourier transform. By setting $\bm{k}=k(\cos\varphi,\sin\varphi)$ and $\bm{R}=R\left(\cos\theta,\sin\theta\right)$, we have $\bm{k}\cdot\bm{R}=kR\cos\big(\varphi-\theta\big)$. We introduce the dimensionless parameter $u=k_F/k_0$ and obtain the following expression for the coordinate-space matrix Green function for $k\approx k_F$:
\bea
\hat{G}(\varepsilon,\bm{R})\approx\int_0^\infty\frac{dk\,k_F}{2\pi}\int_{-\pi}^{+\pi}\frac{d\varphi}{2\pi}\,\frac{e^{\imath kR\cos(\varphi-\theta)}}{\imath\varepsilon-\varepsilon_+(k)}\frac{E_F+\Delta u^2\cos(2\varphi)\tau_1+\Delta u^2\sin(2\varphi)\tau_2+V\big(u^2-1\big)\tau_3}{2E_F}\,.
\eea

\noi We now employ the integral definition of the ordinary Bessel functions of the first kind:
\begin{align}
\imath^nJ_n(z)=\int_{-\pi}^{+\pi}\frac{d\varphi}{2\pi}\,e^{\imath z\cos\varphi}\cos\left(n\varphi\right)
\end{align}

\noi and carry out the integration over the angle entering in the formula for the matrix Green function. This provides:
\begin{align}
\hat{G}(\varepsilon,\bm{R})=\int_0^\infty\frac{dk\,k_F}{2\pi}\,\frac{1}{\imath\varepsilon-\varepsilon_+(k)}\frac{\big[E_F+V\big(u^2-1\big)\tau_3\big]J_0(kR)-\Delta u^2\big[\cos(2\theta)\tau_1+\sin(2\theta)\tau_2\big]J_2(kR)}{2E_F}\,.
\end{align}

\noi The next step is to integrate out the radial wavenumber part. We consider the case where the adatoms distance $R$ is much larger than the Fermi wavelength $\lambda_F=2\pi/k_F$. Given this assumption, we employ the following approximate forms $
J_n(z)\approx\sqrt{2/(\pi z)}\cos\big(z-\pi/4-n\pi/2\big)$ and find the expression:
\begin{align}
\hat{G}(\varepsilon,\bm{R})\approx \sqrt{\frac{2}{\pi k_FR}}\frac{E_F+V\big(u^2-1\big)\tau_3+\Delta u^2\big[\cos(2\theta)\tau_1+\sin(2\theta)\tau_2\big]}{2E_F}\int_0^\infty\frac{dk\,k_F}{2\pi}\,\frac{\cos\left(kR-\frac{\pi}{4}\right)}{\imath\varepsilon-\varepsilon_+(k)}\,.
\end{align}

\noi We evaluate the integral above by employing the linearized form of the dispersion $\varepsilon_+(k)$ and by considering the quasi-classical $E_F\rightarrow\infty$ we write:
\bea
\int_0^\infty\frac{dk\,k_F}{2\pi}\,\frac{\cos\left(kR-\frac{\pi}{4}\right)}{\imath\varepsilon-\varepsilon_+(k)}\approx-\nu_F \int_{-\infty}^{+\infty}\,d\xi\,\frac{\cos\left(k_FR-\frac{\pi}{4}+\frac{\xi R}{\hbar\upsilon_F}\right)}{\xi-\imath\varepsilon}=-\imath{\rm sgn}(\varepsilon)\pi\nu_F e^{\imath{\rm sgn}(\varepsilon)\left(k_FR-\frac{\pi}{4}\right)-\frac{|\varepsilon|R}{\hbar\upsilon_F}},
\eea

\noi where we introduced the density of states at the Fermi level:
\begin{align}
\nu_F=\frac{k_0^2}{4\pi}\frac{E_F}{ \sqrt{(\Delta^2+V^2)E_F^2-(V\Delta)^2}},
\end{align}

\noi along with the Fermi velocity $\upsilon_F=k_F/2\pi\hbar\nu_F$. Hence, we conclude with the final expression of the coordinate-space matrix Green function:
\begin{align}
\hat{G}(\varepsilon,R,\theta)=\frac{-\imath{\rm sgn}(\varepsilon)\pi\nu_F}{\sqrt{2\pi k_FR}} \frac{E_F+V\big(u^2-1\big)\tau_3+\Delta u^2\big[\cos(2\theta)\tau_1+\sin(2\theta)\tau_2\big]}{E_F}\,{\rm Exp}\left[{\imath{\rm sgn}(\varepsilon)\left(k_FR-\frac{\pi}{4}\right)-\frac{|\varepsilon|R}{\hbar\upsilon_F}}\right]\,.
\end{align}

\subsection{III. RKKY and Chiral Susceptibilities}

We now make use of the above result to find the coefficient of the RKKY term. It is straightforward to obtain:
\bea
\chi(R)=-2\int_{-\infty}^{+\infty}\frac{d\varepsilon}{2\pi}\,{\rm tr}\left[\hat{G}(\varepsilon,R,\theta)\hat{G}(\varepsilon,R,\theta+\pi)\right]
=\frac{\nu_F}{\pi R^2}\sin(2k_FR)\,.
\eea

We follow the same approach in order to calculate the chiral susceptibility. For this purpose, we introduce the vectors $\bm{R}_{ij}=R_1 (\cos\theta_1, \sin\theta_1)$, $\bm{R}_{jk}=R_2 (\cos\theta_2, \sin\theta_2)$, and $\bm{R}_{ki}=R_3 (\cos\theta_3, \sin\theta_3)$, with $R_{1,2,3}>0$ and $\theta_{1,2,3}\in[0,2\pi)$. It is important to point out that $\vartheta_{ijk}$ is odd under any odd number of index exchanges. Hence, $\vartheta_{jik}=\vartheta_{kji}=\vartheta_{ikj}=-\vartheta_{ijk}$. At the same time, these exchanges lead to the transformations $\bm{R}_{1,2,3}\mapsto-\bm{R}_{1,3,2}$, $\bm{R}_{1,2,3}\mapsto-\bm{R}_{2,1,3}$, and $\bm{R}_{1,2,3}\mapsto-\bm{R}_{3,2,1}$, respectively. Hence, only antisymmetric terms under these exchanges contribute to $\vartheta_{ijk}$. The definition of $\vartheta_{ijk}$ leads to:
\bea
&&\vartheta_{ijk}=-4\imath\int\frac{d\varepsilon}{2\pi}\,{\rm tr}\Big[\hat{G}(\varepsilon,\bm{R}_{ij})\hat{G}(\varepsilon,\bm{R}_{jk})\hat{G}(\varepsilon,\bm{R}_{ki})\Big]\no\\
&&=-\frac{8(k_F\nu_F)^2}{\sqrt{2\pi}}\frac{\sin\big[k_F(R_1+R_2+R_3)-3\pi/4\big]}{\sqrt{k_F^5R_1R_2R_3(R_1+R_2+R_3)^2}}\left(\frac{k_F}{k_0}\right)^6\left[1-\left(\frac{k_0}{k_F}\right)^2\right]\frac{V\Delta^2}{E_F^3}\sin\big(\theta_1-\theta_2\big)\sin\big(\theta_2-\theta_3\big)\sin\big(\theta_3-\theta_1\big).\no\\
\eea

\noi In the case of an equilateral triangle we can choose $\bm{R}_i=a(1,0)$, $\bm{R}_j=a(-1/2,\sqrt{3}/2)$, and $\bm{R}_k=a(-1/2,-\sqrt{3}/2)$, which yield $R_{1,2,3}=\sqrt{3}a\equiv R$ and $\theta_1=-\pi/6$, $\theta_2=\pi/2$, and $\theta_3=-5\pi/6$. This gives the following angles $\theta_1-\theta_2=\theta_2-\theta_3=\theta_3-\theta_1=-2\pi/3$.

\section{{\large C. Calculations for Chiral Superconductors}}

In this section, we repeat the above procedure for a chiral $d+id$ superconductor. Here, $\mu=0$ and the Fermi level is set by $V=E_F$. We employ the quasi-classical approach also in the upcoming analysis, within which $\Delta\ll E_F$. Therefore, under these conditions, the Hamiltonian of the system takes the approximate form $\hat{H}(k)\approx \xi(k)\tau_3+\Delta\big[\cos(2\varphi)\tau_1+\sin(2\varphi)\tau_2\big]$, with the linearized normal phase dispersion $\xi(k)=\hbar\upsilon_F(k-k_F)$. Note that here $k_F=k_0$, $\hbar\upsilon_F=2E_F/k_F$, while the normal phase density of states at the Fermi level becomes $\nu_F=k_F/2\pi\hbar\upsilon_F=k_F^2/4\pi E_F$.

\subsection{I. Coordinate-Space Matrix Green Function}

Once again we employ the definition of the Green function and write:
\bea
\hat{G}(\varepsilon,\bm{R})&\approx&-\int_0^\infty\frac{dk\,k_F}{2\pi}\int_{-\pi}^{+\pi}\frac{d\varphi}{2\pi}\, e^{\imath kR\cos(\varphi-\theta)}\frac{\imath\varepsilon+\Delta \cos(2\varphi)\tau_1+\Delta \sin(2\varphi)\tau_2+\xi(k)\tau_3}{\xi(k)^2+\Delta^2+\varepsilon^2}\no\\
&=&-\int_0^\infty\frac{dk\,k_F}{2\pi}\frac{\big[\imath\varepsilon+\xi(k)\tau_3\big]J_0(kR)-\Delta \big[\cos(2\theta)\tau_1+\sin(2\theta)\tau_2\big]J_2(kR)}{\xi(k)^2+\Delta^2+\varepsilon^2}\,.
\eea

\noi Also here, we proceed by restricting to the case in which the smallest inter-adatom distance exceeds substantially the Fermi wavelength. Under this assumption, the Bessel functions are approximated with their asymptotic form and we find:
\bea
\hat{G}(\varepsilon,R,\theta)&\approx&-\sqrt{\frac{2}{\pi k_FR}}\int_0^\infty\frac{dk\,k_F}{2\pi}\frac{\imath\varepsilon+\xi(k)\tau_3+\Delta \big[\cos(2\theta)\tau_1+\sin(2\theta)\tau_2\big]}{\xi(k)^2+\Delta^2+\varepsilon^2}\,\cos\left(kR-\frac{\pi}{4}\right)\no\\
&\approx&-\nu_F\sqrt{\frac{2}{\pi k_FR}}\,\cos\left(k_FR-\frac{\pi}{4}\right)\int_{-\infty}^{+\infty}d\xi\,\frac{\imath\varepsilon+\Delta \big[\cos(2\theta)\tau_1+\sin(2\theta)\tau_2\big]}{\xi^2+\Delta^2+\varepsilon^2}\,e^{\frac{\imath\xi R}{\hbar\upsilon_F}}\no\\
&&-\nu_F\sqrt{\frac{2}{\pi k_FR}}\,\sin\left(k_FR-\frac{\pi}{4}\right)\int_{-\infty}^{+\infty}d\xi\,\frac{\imath\xi\tau_3}{\xi^2+\Delta^2+\varepsilon^2}\,\,e^{\frac{\imath\xi R}{\hbar\upsilon_F}}\no\\
&=&-\pi\nu_Fe^{-\frac{\sqrt{\Delta^2+\varepsilon^2}R}{\hbar\upsilon_F}}\sqrt{\frac{2}{\pi k_FR}}\left\{\cos\left(k_FR-\frac{\pi}{4}\right)\frac{\imath\varepsilon+\Delta \big[\cos(2\theta)\tau_1+\sin(2\theta)\tau_2\big]}{\sqrt{\Delta^2+\varepsilon^2}}-\sin\left(k_FR-\frac{\pi}{4}\right)\tau_3\right\}.\qquad
\eea

\subsection{II. RKKY and Chiral Susceptibilities}

We now make use of the above result to find the coefficient of the RKKY term. We remind the reader that, when employing the four component Bogoliubov - de Gennes formalism, one is required to multiply all traces by a factor of $\nicefrac{1}{2}$ so to avoid double counting the electronic degrees of freedom. We thus have:
\bea
\chi(R)=-\int_{-\infty}^{+\infty}\frac{d\varepsilon}{2\pi}\,{\rm tr}\left[\hat{G}(\varepsilon,R,\theta)\hat{G}(\varepsilon,R,\theta+\pi)\right]
&=&
\frac{2\nu_F}{\pi R^2}\int_0^\infty d\varepsilon \,\frac{\varepsilon^2\sin(2k_FR)-(R/\xi_{\rm sc})^2}{\varepsilon^2+(R/\xi_{\rm sc})^2}\,e^{-2\sqrt{\varepsilon^2+(R/\xi_{\rm sc})^2}},
\eea

\noi where we introduced the superconducting coherence length $\xi_{\rm sc}=\hbar\upsilon_F/\Delta$. Here, we are mainly interested in the limit $R\ll\xi_{\rm sc}$. In this case, $|\varepsilon|\gg R/\xi_{\rm sc}$ and we can drop $R/\xi_{\rm sc}$ from the exponential. This approximation yields the result:
\begin{align}
\chi(R)=\frac{\nu_F}{\pi R^2}\left[\sin(2k_FR)-\frac{\pi R}{\xi_{\rm sc}}\right].
\end{align}

\quad We follow once again the same approach and calculate the chiral susceptibility. We define $\bm{R}_{ij}=R_1(\cos\theta_1, \sin\theta_1)$, $\bm{R}_{jk}=R_2 (\cos\theta_2, \sin\theta_2)$, and $\bm{R}_{ki}=R_3 (\cos\theta_3, \sin\theta_3)$, with $R_{1,2,3}>0$ and $\theta_{1,2,3}\in[0,2\pi)$. After accounting for a factor of $\nicefrac{1}{2}$, we pick out the appropriate antisymmetric part and find the expression:
\bea
\vartheta_{ijk}&=&-2\imath\int\frac{d\varepsilon}{2\pi}\,{\rm tr}\Big[\hat{G}(\varepsilon,\bm{R}_{ij})\hat{G}(\varepsilon,\bm{R}_{jk})\hat{G}(\varepsilon,\bm{R}_{ki})\Big]\no\\
&=&-\frac{2(2\pi)^{3/2}\nu_F^3\Delta}{\sqrt{k_F^3R_1R_2R_3}}
\Bigg\{\ph\cos\left(k_FR_1-\frac{\pi}{4}\right)\cos\left(k_FR_2-\frac{\pi}{4}\right)\sin\left(k_FR_3-\frac{\pi}{4}\right)\sin\big[2\big(\theta_1-\theta_2\big)\big]\no\\
&&\,\qquad\quad\qquad\qquad+
\cos\left(k_FR_2-\frac{\pi}{4}\right)\cos\left(k_FR_3-\frac{\pi}{4}\right)\sin\left(k_FR_1-\frac{\pi}{4}\right)\sin\big[2\big(\theta_2-\theta_3\big)\big]\no\\
&&\,\quad\qquad\qquad\qquad+
\cos\left(k_FR_3-\frac{\pi}{4}\right)\cos\left(k_FR_1-\frac{\pi}{4}\right)\sin\left(k_FR_2-\frac{\pi}{4}\right)\sin\big[2\big(\theta_3-\theta_1\big)\big]\Bigg\}\,.
\eea

\end{widetext}

\end{document}